\documentclass[10pt,11pt]{article}
\usepackage{amssymb}
\usepackage{graphicx}
\usepackage{amsmath}
\usepackage{hyphenat}

\input{tcilatex}
\begin{document}

\title{Intraday Patterns in the Cross-section of Stock Returns}
\author{STEVEN L. HESTON, ROBERT A. KORAJCZYK, and RONNIE SADKA\thanks{%
\fontsize{1em}{1em}\selectfont Heston is with the University of Maryland,
Korajczyk is with Northwestern University, and Sadka is with Boston College.
\ We thank Joseph Cerniglia, Ian Domowitz, Lisa Goldberg, Campbell Harvey
(the editor), Ravi Jagannathan, Bruce Lehmann, Maureen O'Hara, Michael
Pagano, Mark Seasholes, Avanidhar Subrahmanyam, Dimitri Vayanos, Julie Wu,
the anonymous referees and associate editor, and seminar participants at the
NBER Microstructure Meeting; Center for Research in Econometric Analysis of
Time Series, Aarhus Universitet; Boston College; Brandeis University;
Cornell University; CREST-INSEE, Paris; HEC Montr\'{e}al; McGill University;
Northwestern University; University College Dublin; University of Southern
California; the Citigroup Quant Conference; Goldman Sachs; and PanAgora
Asset Management for comments. We thank Lew Thorson for computational
assistance. We acknowledge financial support from PanAgora Asset Management
and Korajczyk acknowledges the financial support of the Zell Center for Risk
Research and the Jerome Kenney Fund. Forthcoming in the \textit{Journal of
Finance.}} }
\date{May 3, 2010}
\maketitle

\begin{abstract}
\fontsize{1em}{1em}%
\setcounter{page}{1}%
\noindent Motivated by the literature on investment flows and optimal
trading, we examine intraday predictability in the cross-section of stock
returns. We find a striking pattern of return continuation at half-hour
intervals that are exact multiples of a trading day, and this effect lasts
for at least 40 trading days. Volume, order imbalance, volatility, and
bid-ask spreads exhibit similar patterns, but do not explain the return
patterns. We also show that short-term return reversal is driven by
temporary liquidity imbalances lasting less than an hour and bid-ask bounce.
Timing trades can reduce execution costs by the equivalent of the effective
spread.

\selectfont 
\end{abstract}

\thispagestyle{empty}\bigskip

\bigskip \newpage

\thispagestyle{empty}%
\setcounter{page}{1}%
\noindent We postulate that systematic trading and institutional fund flows
lead to predictable patterns in trading volume and order imbalances among
common stocks. If these patterns are fully anticipated, then they should not
cause predictability in stock returns. Yet, we find periodicity in the
cross-section of stock returns. To study the nature of intraday periodicity
we divide the trading day into 13 half-hour trading intervals. A stock's
return over a given trading interval is negatively related to its returns
over recent intervals, consistent with the negative autocorrelation induced
by bid-ask bounce and lack of resiliency in markets. However, there is a
statistically significant positive relation between a stock's return over a
given interval and its subsequent returns at daily frequencies (i.e., lags
of 13, 26, 39, ... periods). That is, knowing that the equity return of XYZ,
Inc. is high between 1:30 PM and 2:00 PM today has explanatory power for the
return on XYZ, Inc. equity at the same time tomorrow and on subsequent days.
This effect is statistically significant for at least 40 trading days.

Two disparate strands of the literature, when taken together, suggest the
presence of intraday periodicity in trading volume and order imbalances.
First, there is substantial evidence that fund flows to certain types of
institutional investors exhibit autocorrelation. Del Guercio and Tkac (2002,
Table 5) find that flows to mutual funds exhibit significant autocorrelation
while flows to pension fund managers show much weaker autocorrelation.
Frazzini and Lamont (2008, Figure 1) show that fund flows into mutual funds
in the current quarter are related to fund flows in many future quarters (up
to 18 quarters). Lou (2008) also finds evidence of persistent fund flows
into and out of mutual funds. Blackburn, Goetzmann, and Ukhov (2007) find
that the autocorrelation in mutual fund flows varies by category, with value
funds exhibiting greater autocorrelation in flows than growth funds. Even
without explicit autocorrelation in fund flows, there may be autocorrelation
in trading by a given institutional manager. For example, some managers
rebalance individual customers' separate accounts on successive days.
Campbell, Ramadorai, and Schwartz (2009) use quarterly 13-F filings and
intraday trading data from the New York Stock Exchange's Trade and Quotation
(TAQ) database to estimate the aggregate trading behavior of institutional
investors at a daily frequency. Their evidence indicates that institutional
trading is highly persistent. That is, institutional investment managers
tend to buy or sell the same stocks on successive days.

Second, there is a growing literature on trading algorithms that are
designed to minimize trading costs or optimize the trade-off between trading
costs and the uncertainty associated with the prices obtained over the
trading program. For example, Bertsimas and Lo (1998) study the problem of
minimizing the expected cost of executing an exogenously specified trade
size over an exogenously given horizon. In the simplest version of their
model the price impact of trade is permanent and proportional to the number
of shares traded while non-trade public information causes mean zero asset
price changes. The lowest expected impact strategy is to trade an equal
number of shares at every point in time. If conditional mean price changes
are nonzero, then it would be optimal to either speed up or slow down
trading relative to this base case (depending on the expected direction of
price movements and the direction of trade).

A number of papers extend the above results to incorporate the risk of the
trading program, in addition to the expected trading cost, for risk averse
traders (e.g., Almgren and Chriss (2000), Grinold and Kahn (2000), Vayanos
(2001), Huberman and Stanzl (2005), and Engle and Ferstenberg (2007)).\
Under price dynamics similar to Bertsimas and Lo (1998) and a price impact
function with permanent and transitory components, Almgren and Chriss (2000)
derive an efficient frontier of optimal trading strategies that explicitly
trade off execution costs with the risk of adverse price movements when
traders have mean-variance utility. Linear price impact functions allow for
closed-form solutions. The expected cost-minimizing strategy of Bertsimas
and Lo (1998) is optimal for risk-neutral agents. Risk averse agents will
generally liquidate the portfolio more rapidly in order to reduce the
variance of the execution cost. Thus, these papers imply a trading
trajectory that executes quickly at the beginning of the trading horizon and
slows down as time passes. Hora (2006) uses a different preference
specification that leads to an optimal trading strategy that trades rapidly
at the beginning and end of the trading horizon and more slowly in the
middle of the period. The horizon over which trading takes place is
generally unspecified in the models and could be a day, month, year, or some
other interval. Almgren and Lorenz (2006) argue that a trading horizon of
one day is consistent with the manner in which many institutions trade. In
line with this view, Breen, Hodrick, and Korajczyk (2002) find that 92.5\%
of a sample of institutional orders are completed the same day that trading
is initiated. The use of algorithmic trading has grown considerably and, as
of 2007, was estimated to account for one-third of equity trading volume in
the United States (Hendershott, Jones, and Menkveld (2009)).

The autocorrelation of fund flows and the use of trading algorithms
discussed above imply the existence of daily patterns in volume and order
imbalances in individual stocks. If Fund A receives fund flows today and
buys stock XYZ using the Almgren and Chriss (2000) algorithm, then A will
trade XYZ more aggressively in the morning and less aggressively as the day
progresses. Since fund flows are autocorrelated, Fund A is likely to receive
fund flows tomorrow. Further, because Fund A is likely to buy XYZ tomorrow
using the same trading algorithm, volume and order imbalances in XYZ are
likely to arise tomorrow at the same time as today. Anecdotal evidence
indicates that this periodicity may be likely to happen even without
explicitly invoking trading algorithms. Large equity managers have demanding
schedules that involve client meetings, staff meetings, compliance, and
reporting. The decision to buy or sell stock may require research analysis
or portfolio optimization. The manager must then decide whether to use
brokers or whether to relay the transactions through a crossing network such
as Posit, Island, or Archipelago. These networks cross at certain times of
the day, leaving unfilled orders at the end of the crossing. A busy manager
may therefore find it cheap and expedient to execute these orders at
specific times of the day, leaving the remaining time available for other
trading, research, and risk management activity. The use of a trading
algorithm that specifies a particular pattern of trades would only
strengthen this periodicity. Additionally, some types of institutional
traders may have other reasons for demonstrating periodicity in trading.
Index funds, for example, tend to trade at the close to reduce tracking
error.

The above conjectures about trading behavior imply periodicity in volume and
order imbalance. Indeed, there is extensive evidence of intraday patterns in
volume and volatility (e.g., Wood, McInish, and Ord (1985), Harris (1986),
Jain and Joh (1988), and Pagano, Peng, and Schwartz (2008)). However, while
intraday patterns of volume and volatility can be justified with models of
discretionary liquidity trading (e.g., Admati and Pfleiderer (1988) and Hora
(2006)), predictable patterns in returns are harder to explain. We consider
several possibilities, none of which fully explain the periodicity. For
instance, a long-standing literature on intraday return patterns (e.g.,
Wood, McInish, and Ord (1985), Harris (1986), and Jain and Joh (1988)) shows
that average returns tend to be higher at the beginning and end of the
trading day. Our observed patterns are not consistent with a simple mean
shift at the beginning and end of the day. Additionally, the literature
documents a number of seasonal return patterns, say at the monthly or
quarterly frequency, in stock returns. Some of this periodicity is
consistent with predictable patterns of trading by investors. For example,
Keim (1989) finds that turn-of-the-year trading patterns induce patterns in
the probability that trades occur at the ask price versus the bid price. He
finds that this trading pattern explains the size-related turn-of-the-year
effect in stock prices. However, we find that the return continuation at
daily intervals is not due to the effect found in Keim (1989), that is, to
changes in the frequency of trades occurring at either the bid or ask
prices. Our intraday return pattern is stronger over the first and last
half-hours of the trading day, as one might expect given the patterns in
volume and volatility, but remains statistically significant over the other
periods of the day. Thus, the intraday return pattern is not due simply to
uniformly high returns at the beginning and end of the trading day.

Markets may also respond to news that arrives on a daily schedule. Newscasts
and conference calls occur at regular times of the day, and behavioral
biases may cause traders to overreact or underreact to salient news (Tversky
and Kahneman (1974)), leading in turn to return periodicity. Our observed
return periodicity is consistent with the information acquisition model of
Holden and Subrahmanyam (2002) if agents tend to become informed at
intervals that correspond to our observed periodicity.

Further, if there are regular news cycles, then traders may be reluctant to
hold stocks during risky times. For example, if market-wide information is
released at 3:30 PM, then high beta stocks may have a higher return premium
at those times. This suggests that daily phenomena will be more pronounced
among risky stocks than safe stocks. We do not find support for this
hypothesis.

We also study whether the return periodicity is driven by similar patterns
in order flow and liquidity. We show that trading volume has similar
patterns to returns, that is, firms that experience a relatively high change
in their trading volume over a particular half-hour interval typically
experience a high change in their volume during the same half-hour interval
during each of the next few days. Although related, the periodicity in
trading volume does not completely explain the periodicity of returns. When
we split volume into that due to large versus small trade size, both
measures of volume show daily periodicity, but neither explains the daily
return periodicity. Oddly, the level of order imbalance (OI) does not
exhibit obvious periodicity, even when partitioned into small versus large
trades. However, changes in order imbalance exhibit periodicity similar to
returns, but the pattern in order imbalances does not explain the pattern in
returns. Bid-ask spreads exhibit similar intraday periodicity as that of
returns, but spreads do not explain the return patterns.

Several other tests suggest that the intraday periodicity at the daily
frequency is not simply an artifact of previously shown patterns. For
example, it is not concentrated in any particular weekday, and therefore is
not a manifestation of the day-of-the-week effect (see French (1980)), and
it is not concentrated in any particular month (see Heston and Sadka (2008a,
2008b)). The effect is also not particularly related to the
turn-of-the-month effect (Ariel (1987)) or the turn-of-the-quarter effect
(Carhart et al. (2002)). The pattern of intraday returns is highly
persistent, lasting for two months (40 trading days). It is not due to a
particular market capitalization group, inclusion in the S\&P 500 index, nor
is it a manifestation of intraday movements in systematic risk. There is
substantial evidence of intraday periodicity of return volatility (Andersen
and Bollerslev (1997)), which we also find in our sample. However, patterns
in volatility do not explain our return periodicity.

Finally, we compare the results for the post-decimalization period to prior
periods, which include 1993 to 1997, for which the minimum price increment
was one-eighth of a dollar, and 1997 to 2000, for which the minimum price
increment was one-sixteenth of a dollar. We find that the statistical
significance of the intraday periodicity is greatest in the
post-decimalization period.

In Section I we present our main findings regarding return continuation at
daily frequencies and short-term return reversals. In Section II we study
possible cross-sectional explanations for the findings. In Section III we
study the properties of volume, order imbalance, volatility, and liquidity
while we provide additional diagnostics in Section IV. Our conclusions are
presented in Section V.

\begin{center}
\textbf{I. Patterns of Resilience in Intraday Stock Returns}
\end{center}

We begin this study by measuring intraday persistence in the cross-section
of stock returns. It is well known that short-term stock returns are
negatively autocorrelated (e.g., Lehmann (1990) and Lo and MacKinlay
(1990)). While this phenomenon does not occur in the model of Glosten and
Milgrom (1985), in which the spreads are due solely to adverse selection
caused by informed traders, it appears in other models with bid-ask spreads
(Roll (1984) and Glosten and Harris (1988)), specialist inventory effects
(Stoll (1978)), or other non-adverse selection costs associated with
market-making. Here we study the resilience of stock prices based on the
pattern of autocorrelation over various horizons.

\noindent \textit{A. Methodology and Main Results}

Our sample of firms consists of all New York Stock Exchange (NYSE) listed
firms from January 2001 through December 2005 that we are able to match with
the NYSE Trade and Quotation (TAQ) database. We retain only assets whose
CRSP share codes are 10 or 11, that is, we discard certificates, ADRs,
shares of beneficial interest, units, companies incorporated outside the
U.S., Americus Trust components, closed-end funds, preferred stocks, and
REITs. The matching yielded data for 1,715 firms. The period of study is
chosen to substantially overlap the period of decimalization, the transition
to which occurred between August 2000 and January 2001. We use the TAQ
database to calculate intraday stock returns. For each stock we calculate
returns over half-hour intervals. This gives 13 intraday intervals per
trading day from 9:30 a.m. to 4:00 p.m. This excludes after-hours trading
and overnight open-close price movements. Note that settlement on stock
transactions occurs after the end of the trading day. This means that trades
at different times do not need to earn the risk-free rate intraday. In other
words, intraday stock returns compensate for liquidity and risk, not for the
time value of money. In addition to returns, we also measure changes in
volume defined as the change in the logarithm of the number of shares traded
over a half-hour interval. The return and volume data provide measures of
price and quantity movements of individual stocks throughout the day.

We analyze intraday stock returns using the cross-sectional regression
methodology of Jegadeesh (1990). For each lag, $k,$ we run cross-sectional
regressions of half-hour stock returns on returns lagged by $k$ half-hour
periods,%
\begin{equation}
r_{i,t}=\alpha _{k,t}+\gamma _{k,t}r_{i,t-k}+u_{i,t},  \label{k lags}
\end{equation}%
where $r_{i,t}$ is the return on stock $i$ in the half-hour interval $t$.
The slope coefficients $\gamma _{k,t}$ represent the response of returns at
half-hour $t$ to returns over a previous interval lagged by $k$ half-hour
periods, the \textquotedblleft return responses.\textquotedblright $^{1}$ In
addition to the simple regression (\ref{k lags}), we also use a multiple
regression to estimate all return responses up to five trading days jointly%
\begin{equation}
r_{i,t}=\alpha _{t}+\gamma _{1,t}r_{i,t-1}+\gamma _{2,t}r_{i,t-2}+...+\gamma
_{65,t}r_{i,t-65}+u_{i,t}.  \label{all lags}
\end{equation}%
Both the simple regression and the multiple regression use all firms with
returns available in intervals $t$ and $t-k$. Initially we present
unconditional return responses, averaging over different times of the day $t$
for each lag $k$. Later we examine conditional averages, including
time-of-day variation in the magnitude of these effects. We calculate the
pattern of return effects by averaging return responses over time for
half-hour lags $k$. Note that using cross-sectional regression in this way
is different from measuring the autocorrelation of stock returns. In
particular, the cross-sectional regression subtracts an overall market
effect, which reduces variance and focuses on returns relative to other
stocks.

Figure 1 presents the average return responses across different lags (Panel
A) for lags up to one week.$^{2}$ The data are scaled so the units for $%
\widehat{\gamma }_{k}$ are basis points. With 13 half-hour intervals per day
and five trading days per week, this produces 65 lagged intervals.
Consistent with the previous literature, the first several return responses
are negative. This means that stock returns experience a reversal period
lasting several hours. These reversals may be a manifestation of bid-ask
bounce, time variation in the frequency of trades occurring at bid versus
ask prices (as in Keim (1989)), or temporary liquidity dislocations that
take some time for liquidity providers to accommodate. We disentangle these
effects.

\begin{center}
[Figure 1 here]
\end{center}

The most striking pattern in Figure 1 is that, following the reversal
period, the return responses are positive, peaking at horizons that are
exact multiples of 13 half-hours, or one trading day. The figure shows clear
periodicity at daily intervals. Panel B of Figure 1 plots the $t$-statistics
of $\widehat{\gamma }_{k}$, as estimated in Fama and MacBeth (1973). The $t$%
-statistics show the same type of periodicity. Over the first week the
smallest $t$-statistic at the daily frequency (lags 13, 26, 39, 52, and 65)
is 9.62.

Table I shows that the simple regression return responses are highly
statistically significant at almost all lags. The Internet Appendix (Table
IA.I)$^{3}$ shows the results of the multiple regression responses are
similar to the simple regression in magnitude, generally with greater
statistical significance. Over the period of one calendar day these results
indicate that returns are temporarily reversed but then rebound.

\begin{center}
[Table I here]
\end{center}

Looking beyond 13 lags, the return effects over half-hour intervals on
subsequent days remain largely negative, with statistically significant
positive effects at multiples of 13 lags, that is, 26, 39, 52, and 65. It
appears that temporary price pressure is reversed at virtually all future
times except at the same time interval on subsequent days.

Since the simple regression produces results almost identical to the
multiple regression, the univariate specification (\ref{k lags}) seems
adequate. By focusing on one lag at a time, we can easily examine a number
of different lags.

The implicit portfolio weights in the portfolio returns obtained from the
estimated regression coefficients, $\widehat{\gamma }_{k}$, could be rather
extreme since, as mentioned above, $\widehat{\gamma }_{k}$ is the return on
a zero-cost portfolio whose excess return $k$ periods ago is 100\% (Lehmann
and Modest (2005)). Therefore, we also study the returns to relatively
well-diversified (equal-weighted) long-short portfolios formed on the basis
of lagged half-hour returns. Using the methodology of Jegadeesh and Titman
(1993), we sort stocks into equal-weighted deciles based on their returns
over a previous half-hour interval. Figure 2 and Table II present these
results extending the lags to roughly two trading months (520 lags). We form
portfolios of "losers" and "winners" and calculate equal-weighted returns
(in basis points) on portfolios of stocks that had the lowest and highest
10\% of returns $k$ periods ago, respectively. The column labeled "10-1" is
the average return on a portfolio that is long the winners portfolio and
short the losers portfolio.$^{4}$ This long-short average return is plotted
as a function of $k$ in Figure 2, along with the associated $t$-statistics.

\begin{center}
[Figure 2 and Table II here]
\end{center}

Based on a half-hour return on a given day, the average difference between
the top decile of winners and the bottom decile of losers is 3.01 basis
points at the same time on the next day. To get a sense of the economic
magnitude of this periodicity, we consider a few comparisons. Being able to
capture three basis points is similar in magnitude to the daily equity risk
premium.$^{5}$ The effect may be larger than the equity risk premium in
terms of incremental Sharpe ratio since one needs to bear one day of risk to
earn the risk premium but only needs to shift trade a fraction of a day to
take advantage of the periodicity. Being able to capture three basis points
is also similar in magnitude to saving a \$0.01 commission on a stock priced
at \$30/share (which is in line with institutional commission rates on
common stock). Alternatively, Gurliacci, Jeria, and Sofianos (2009) find
that the typical quoted half-spread of S\&P 500 stocks in their benchmark
("normal" volatility) period (May 2008) is 1.7 basis points, which means
that the average return predictability is of the same order of magnitude as
twice the quoted half-spread (the return difference requires two trades, so
the relevant comparison is twice the half spread).$^{6,7}$ Jeria and
Sofianos (2008) evaluate institutional order executions for two alternative
algorithms and find that there is a 2.3 basis point difference between the
implementation shortfall for the average order. They use this result to
recommend that traders migrate from one algorithm to the other. Thus, for
high-frequency institutional traders, the effect appears to be economically
consequential, particularly for high frequency traders ("Rise of the
Machines," \textit{The Economist,} July 30, 2009).

The average return difference remains positive, albeit smaller, on
subsequent days. The difference remains positive and statistically
significant for up to 40 days (520 half-hours). It appears there is a
persistent and predictable pattern in intraday stock transaction prices.
When a stock goes up on one day, buyers earn a return premium by buying the
stock prior to (or delaying sales until after) the same time interval on
future days. Conversely, sellers provide a discount by selling at this time
when they could expect a higher average price 30 minutes later.

\noindent \textit{B. Reversals: Resiliency, Bid-Ask Bounce, and Periodicity
in Order Imbalances}

A market is resilient when it replenishes depth rapidly after having been
depleted by large transactions (see, for example, Coppejans, Domowitz, and
Madhavan (2004) and Large (2007)). We expect to see large price movements
due to these types of trades, followed by rapid reversals as new limit
orders replenish the order book. In less resilient markets, the reversal
will take longer, leading to a longer period of return reversals. Therefore,
the observed negative values of $\widehat{\gamma }_{k}$ at low values of $k$
(Figure 1, Table I, and the Internet Appendix Table IA.I), and the negative
returns on the winners minus losers portfolios (Figure 2) may be due to
liquidity shocks when the market is not very resilient. While this cause may
be plausible for the negative $\widehat{\gamma }_{k}$ for $k$ between one
and eight, it is unlikely to explain the negative $\widehat{\gamma }_{k}$
for lags corresponding to two to five days ago.

An alternative explanation for the negative values of $\widehat{\gamma }_{k}$%
\ is some combination of bid-ask bounce and time variation in the frequency
of trades occurring at bid versus ask prices (as in Keim (1989)). In Roll
(1984) there is an equal chance that a trade occurs at the bid price or the
ask price. This implies that order flow is not autocorrelated, but that
bid-ask bounce induces negative first-order autocorrelation in returns. In
the Roll model there is no autocorrelation in returns at lags greater than $%
k=1$. However, the empirical evidence shows that order imbalances (signed
order flow) tend to be autocorrelated (Sadka (2006)), so trades at the ask
will tend to be followed by trades with a higher probability of being at the
ask. This type of pattern could lead to reversals that are not immediate
(that is, return autocorrelation beyond the first lag).

Of course, both liquidity imbalances and bid-ask bounce are likely to be
present in the data. To distinguish between lack of resiliency and bid-ask
bounce, we reestimate $\widehat{\gamma }_{k}$ using returns calculated from
only bid prices (defined as the current bid quote for the transactions),
from only ask prices, and from only midpoint prices, instead of from
transaction prices. If the observed return reversal pattern is due to lack
of resiliency, we should see return reversals using the bid-to-bid,
ask-to-ask, and midpoint-to-midpoint returns. Conversely, if the reversals
are merely due to bid-ask bounce, then there should be no return reversals
using bid-to-bid or ask-to-ask returns.

In Figure 3 we plot the estimates, $\widehat{\gamma }_{k},$ and their $t$%
-statistics using bid-to-bid, ask-to-ask, and midpoint-to-midpoint returns.
The only reversal that is statistically significant is that for $k=1$. This
is consistent with the market taking longer than 30 minutes, but less that
60 minutes, to restore shocks to liquidity, consistent with the evidence in
Chordia, Roll, and Subrahmanyam (2005). Thus, there is evidence of a
resiliency effect at lag $k=1$ only, and all of the reversals evident using
transaction returns beyond lag $k=1$ disappear using bid-to-bid and
ask-to-ask returns. Our inference is that the negative values of $\widehat{%
\gamma }_{k}$ beyond lag 1 shown in Figures 1 and 2 are due to bid-ask
bounce. Once that effect is removed, there seems to be some evidence of
momentum (positive values of $\widehat{\gamma }_{k}$) in returns using
bid-to-bid and ask-to-ask prices.

\begin{center}
[Figure 3 here]
\end{center}

While the short-term reversals evident in Figures 1 and 2 are consistent
with microstructure theory and give us a coherent picture of resiliency and
bid-ask bounce, the positive return responses at daily frequencies are much
more difficult to explain. We now turn to the nature of these return
continuations and their relation to other periodicity anomalies. We first
study patterns across past-return deciles and across different times of the
day.

\noindent \textit{C. Patterns Across Past-Return Deciles}

We develop and compare strategies that hold stocks over a single half hour
interval to focus on the periodic return pattern based on past returns at
different lags. Table III shows the performance of stock deciles ranked on
their performance in previous half-hour intervals. The strategies labeled
"daily" sort stocks based on one half-hour interval at the same time on a
previous day ($k=13,26,39,52,$ or $65$). The top row lists the decile
number, with decile $j$ at time $t$ being made up of those stocks whose
returns are in the $j^{th}$ decile (from lowest to highest) in period $t-k$
for $k=13,26,39,52,$ or $65$.

\begin{center}
[Table III here]
\end{center}

The nondaily strategies rank assets into deciles based on average returns
over a previous 24-hour period except the daily lags. For example, the Day
1-Nondaily strategy forms portfolios on the basis of their average returns
over lags 1 through 12, the Day 2-Nondaily strategy uses average returns
over lags 14 through 25, etc. By studying the returns on these decile
portfolios we can observe any nonlinearity in the relation between average
returns across deciles, such as whether the daily pattern is concentrated in
the upper or lower performing deciles.

The table shows most of the statistical and economic significance comes from
the extreme deciles. For example, the worst nondaily losers over the
previous day earn an average of 3.16 basis points, while the best nondaily
winners lose 1.51 basis points per half-hour holding period. The average
returns are nearly monotonic across intermediate deciles. The signs are
reversed for the daily decile strategies. For example, the worst decile of
losers over the same interval of the previous day continue to lose an
average of 1.35 basis points, while the best decile of daily winners earn
1.66 basis points per half-hour interval. The intermediate decile average
returns are monotonic, but most of the significance comes from the lowest
and highest deciles.

The magnitude of the daily response is most substantial on the first day,
however there is a smaller persistent effect. For lags beyond one day, the
average daily winners minus losers decile spreads remain above one basis
point per half-hour for lags of up to five days (with $t$-statistics above
8.7). By contrast, the magnitude of the negative nondaily winners minus
losers decile spreads is less than one basis point. Average decile spreads
for nondaily strategies are statistically significant for lags of at least
four days.

\noindent \textit{D. Time of Day}

The return patterns in Figures 1 and 2 reflect average behavior throughout
the day. Investors can pay up to three extra basis points, on average, to
execute trades at a particular time of day. However, it is possible this
effect is concentrated at certain times of the day. We first investigate
whether the intraday pattern is an artifact of biases in opening or closing
prices. Overnight orders are executed at the open, and many traders (for
example, index funds concerned with minimizing daily tracking error) place
market-on-close orders. Temporary price distortions caused by opening and
closing procedures might produce predictability in stock returns that does
not affect stock prices at other times of the day.

Table IV shows the excess return of winners minus losers decile spread
strategies during different half-hour intervals throughout the day. Since
the nondaily effect is mainly driven by bid-ask bounce, we only report the
results for the daily strategy. The full results are reported in the
Internet Appendix (Table IA.III). For Day 1, the daily decile spreads are
sorted based on the returns for lag 13, while the nondaily spreads (in the
Internet Appendix) are sorted based on the average returns for lags one
through 12. The return effect is quite pronounced in the first and last
half-hours of trading. The Day 1 daily decile spread earns over 11 basis
points in the opening half-hour, while the Day 1 nondaily strategy loses
over eight basis points. This is a difference of 19 basis points between
these strategies in the opening half-hour. Similarly, the Day 1 daily decile
spread earns over eight basis points near the close of trading, while the
corresponding nondaily strategy loses 11 basis points. A smaller effect
remains during the middle of the day. The Day 1 daily strategy earns
positive average excess returns in every half-hour interval from 10:00 a.m.
to 3:30 p.m., averaging 1.75 basis points over this period. Meanwhile, the
Day 1 nondaily strategy declines in nearly every half-hour period, losing
3.74 basis points per half-hour over this period. This is a consistent
pattern throughout the day. At longer lags, labeled Day 2 through Day 5 in
Table IV, the return patterns are of a similar shape but smaller in absolute
value. When we average the second through twelfth intervals (that is,
average returns from 10:00 a.m. to 3:30 p.m.), all the daily spreads are
significantly different from zero at conventional levels while the nondaily
spreads are statistically significant for Day 1 and Day 2.

\begin{center}
[Table IV here]
\end{center}

Figure 4 shows that the periodic daily pattern of return strategies has a
similar shape but different magnitudes at different times of the day. The
daily strategies earn a premium in the first half-hour of trading, the last
half-hour, and in the intermediate intervals from 10:00 to 3:30. This
evidence indicates that while the periodicity at daily intervals is stronger
at the beginning and end of the trading day (as can be seen from the
magnitudes of the coefficients in the left side of the figure), it exists at
all times (as can be seen from the magnitudes of the \textit{t}-statistics
in the right side of the figure).$^{8}$ For the intermediate intervals there
is more anticipation of the periodicity at lag 13 since the coefficients are
significantly positive for lags 10 through 13. The top and bottom graphs in
Figure 4 show some leakage from the closing half-hour to the next day's
opening half-hour. Figure 4 is consistent with the argument that the
patterns are mainly, although not exclusively, driven by trading at or near
the opening or closing of the trading day.

\begin{center}
[Figure 4 here]

\textbf{II. Potential Cross-sectional Explanations}
\end{center}

The pattern of daily return continuation has several potential explanations.
One concern is the effect of data errors caused by thinly traded stocks.
After addressing this, we consider whether size or risk can explain the
return pattern. We also examine whether trading profits of strategies based
on return periodicity are wiped out by paying the bid-ask spread.

\noindent \textit{A. Low Price and Thin Trading}

A major concern is whether the return continuation is driven by thinly
traded stocks with inaccurately or irregularly measured returns. Returns on
low-priced stocks are most strongly affected by discreteness. However, since
our sample is from the post-decimalization period, we do not expect
discreteness to be a major cause of return periodicity.

Many stocks may not be traded throughout a half-hour interval. For example,
a stock might trade once at 12:01, and then a second time at 12:02, without
any further activity before 12:30. While it is not clear whether this would
impart a daily pattern in measured returns, we verify whether our results
are contaminated by such inaccurate or illiquid returns.

Table V addresses these issues by repeating the decile spread analysis of
Table III while eliminating low-priced and thinly traded stocks. We define
low-priced stocks as stocks with share prices below \$5. Removing the
low-priced stocks decreases the nondaily reversal effect (reported in
Internet Appendix Table IA.IV). The average nondaily decile spread on Day 1
changes from -4.67 basis points per half-hour in Table III to -1.77 basis
points (with a $t$-statistic of -8.07) in the Internet Appendix (Table
IA.IV). Given that the reversals beyond lag $k=1$ are driven by bid-ask
bounce, one would expect some attenuation since low-priced and thinly traded
stock tend to have higher percentage spreads.

\begin{center}
[Table V here]
\end{center}

The daily continuation effect shrinks less than the nondaily reversal
effect. Table V shows that the daily decile spread earns 1.93 basis points
(with a $t$-statistic of 10.34) on Day 1, whereas it earns roughly three
basis points with all stocks in Table III. In both Table III and Table V the
daily decile spread remains roughly one basis point per half-hour on Days 2
through 5.

To address problems of thin trading, we also eliminate stocks that do not
have an average of at least 10 trades per half-hour interval over the
previous month. In other words, we use stocks that average at least one
trade every three minutes. This retains roughly 80\% of the stocks in our
sample. The second column of Table V shows that this restriction does not
affect the results as much as eliminating the low-priced stocks. The daily
decile spread earns 2.45 basis points (with a $t$-statistic of 18.44) per
half-hour on Day 1, while the nondaily decile spread (in the Internet
Appendix (Table IA.IV)) loses 3.17 basis points (with a $t$-statistic of
-19.44). Results on subsequent days are similar to returns using the full
sample, reported in Table III. Overall, the reversal and continuation
patterns are not an artifact of mismeasurement of the return on low-priced
stocks or mistiming of returns on thinly traded stocks.

\noindent \textit{B. Size and Transactions Costs}

A concern about the nature of intraday patterns in stock returns is market
depth. If there is a return premium at certain times of the day then it may
be compensation for illiquidity that makes stocks difficult to trade at
efficient prices during those times. For example, Admati and Pfleiderer
(1988) develop a model where traders pool trades in certain periods of the
day in order to take advantage of the increased liquidity of pooling. This
implies a pattern primarily among smaller and less liquid stocks that face
larger adverse selection problems. To the extent that firm size proxies for
variation in adverse selection problems, it provides a control variable.

In Table VI we sort stocks into three groups based on market capitalization
at the end of the previous calendar year. In particular, the table reports
the decile spread strategies separately for small-, medium-, and
large-capitalization firms, where each group contains 33\% of the sample.
The nondaily strategies on Day 1 (reported in the Internet Appendix (Table
IA.V)) have returns that are more negative among small firms. This is
consistent with small-cap firms having larger proportional spreads. The Day
1 nondaily decile spread loses more than 10 basis points in the opening
half-hour and more than 22 basis points in the closing half-hour, while
averaging a loss of more than eight basis points in the mid-day half-hour
intervals. Conversely, the daily decile spread strategies are more
profitable with small stocks. The Day 1 daily strategy averages more than
five basis points per half-hour. In contrast, average returns are in the
range of one to three basis points for medium and large stocks. The average
excess returns for strategies based on longer daily and nondaily lags are
smaller and do not differ much across size categories. However, almost all
daily strategies maintain statistical significance at the 95\% level in all
size categories at the open, mid-day, and close. This indicates that while a
liquidity/microstructure effect explanation may have merit, return
periodicity is not associated exclusively with small firms.

\begin{center}
[Table VI here]
\end{center}

An important consideration is the magnitude of transaction costs associated
with the trading strategies implicit in our reported returns. We find
predictable excess returns of several basis points within a half-hour
interval based on transaction prices. However, a trader with no other motive
for trade must pay the ask price or accept the offer price to get immediate
execution and avert time slippage. Larger orders also lead to larger price
impacts. Table VII reports the decile spread results for strategies that buy
at the offer and sell at the bid.\textbf{\ }The average results are negative
for all size categories at all times of the day, indicating that the
periodicity we find does not indicate a pure profit opportunity. It is
important to remember that these results involve the round-trip transaction
costs of two different long-short decile strategies. Therefore, they incur
the average cost of four transactions. For example, among small stocks the
average decile spread return on the Day 1 Nondaily strategy (reported in the
Internet Appendix (Table IA.VI)) is -25.03 basis points, and the average
decile spread return on the Day 1 Daily strategy is -23.78 basis points. The
Internet Appendix (Table IA.V) shows that the Day 1 Nondaily strategy lost
9.92 basis points among small stocks while Table VI shows that the Day 1
Daily strategy gained 5.16 basis points. Comparing the pre-transaction cost
returns in Table VI to the post-spread return in Table VII, we see that the
implied one-way spread cost is between four and seven basis points. While
the difference between the performance of Daily and Nondaily strategies does
not cover the costs of two round-trips, it often compares favorably with the
magnitude of one-way transaction costs. This suggests that many investors
have a demand for immediate execution of trades and are not willing to shift
their trades by 30 minutes to take advantage of the periodicity.

\begin{center}
[Table VII here]
\end{center}

These trading strategies are not likely to be profitable for stocks with
large bid-ask spreads. Therefore, the results in Table VII exclude stocks
that have a quoted relative spread of more than 10 basis points at the
beginning of a given trading interval. We also perform the analysis
restricted to stocks with spreads less than five and 25 basis points,
respectively. The results are similar to those in Table VII. Note that very
few small stocks have spreads less than five basis points, so there is no
point conditioning on less than five basis points. Since our reported raw
profits are not greater than 25 basis points, we do not condition on spreads
greater than 25 basis points. The results suggest that it is quite difficult
to profit from these type of intraday strategies without some exogenous
desire to trade.

The transaction costs for medium and large stocks are substantially smaller
than those for small stocks. Table VII shows that for medium stocks the
average decile spread results are roughly -20 basis points and for large
stocks they are roughly -14 basis points. This corresponds to one-way
trading costs of less than five basis points. The losses\textbf{\ }of the
Nondaily and Daily strategies in Table VII and the Internet Appendix (Table
IA.VI) are smaller for medium and large stocks than for small stocks.

Our evidence for the nondaily reversal effect is consistent with the results
of Avramov, Chordia, and Goyal (2006), who find that the reversals at weekly
and monthly intervals are smaller than transactions costs.

\noindent \textit{C. Beta}

If high frequency changes in risk or liquidity influence investors' demand
for assets, then we might find periodicity in returns due to periodicity in
risk or liquidity. For example, stocks might be riskier at certain times of
the day when news is released, or they might be subject to institutional
transactions that follow a daily cycle. This section explores these
possibilities.

A priori, it seems unlikely for stocks to have large fluctuations in
systematic risks during the day because companies do not change their
financial exposures from hour to hour. On the other hand, some economic
series are released at scheduled times, and firms may have exposure to
systematic news released at those times. In this case traders may be
reluctant to hold stocks at these risky times. To assess this possibility we
control for risk by regressing stock returns on the equal-weighted market
index. To correct for nonsynchronous trading (Dimson (1979)), we include the
contemporaneous half-hour market return along with 13 leads and lags. The
Internet Appendix (Table IA.VII) reports the average intercepts from these
regressions. Since the intraday interest rate is effectively zero, these
intercepts have the interpretation of risk-adjusted returns.

The results reported in the Internet Appendix (Table IA.VII) resemble the
previous results using unadjusted average returns. The average risk-adjusted
return on decile return spreads of lag 13 winners in excess of lag 13 losers
is 3.03 basis points. The decile spread has a risk-adjusted return of -4.65
basis points when portfolios are formed on the basis of returns over lags 1
through 12. These effects are particularly pronounced in the first half-hour
and last half-hour of the day, but remain statistically significant in the
middle of the day.

The average risk-adjusted returns for the daily decile spreads continue to
be substantial in the opening and closing half-hour even when sorting on
half-hour returns up to five business days ago. The Day 5 average decile
spread for $k=65$ is 4.84 basis points in the first half-hour, and 3.42
basis points in the last half-hour. The effect is much smaller in the middle
of the day, less than one basis point, but remains statistically
significant. Therefore, controlling for market risk does not eliminate the
daily pattern.

These results also show that the periodicity is not an artifact of all
stocks having high returns at certain times of the day and low returns at
others, that is, simple mean shifts across half-hour intervals. If this were
driving the results, risk-adjusted returns should eliminate the effect since
the risk adjustment takes out the market return. However, the risk-adjusted
returns in the Internet Appendix (Table IA.VII) are almost identical to the
unadjusted returns in Table IV.

\begin{center}
\textbf{III. Volume, Order Imbalance, and Related Variables}
\end{center}

Autocorrelated fund flows and a preference for trades at particular times of
the day should lead to periodicity in quantity variables (e.g., volume) and
signed quantities (e.g., order imbalance), and possibly also in volatility
and liquidity. As discussed above, at the aggregate level there is an
extensive literature on intraday patterns in volume and volatility. This
literature is consistent with periodicity in volume, order imbalance,
volatility, and liquidity of the type observed in returns. However, there
may be different intraday patterns that are also consistent with this
literature (for example, an intraday step function in which the means of all
variables are shifted each period). We wish to determine whether these
variables have periodicity similar to returns. If they do, we would like to
know if past observations of these variables explain the return periodicity.
We find common periodic daily patterns in some of these variables, but show
that these variables do not subsume the pattern of return predictability.

All these variables are quite persistent in their levels, so the response
coefficients we observe by estimating equation (\ref{k lags}) look different
than those for returns in Figure 1. This is to be expected, given what we
know about volume (Lo and Wang (2006)), order imbalance (Sadka (2006)),
volatility (Bollerslev (1986)), and liquidity (Korajczyk and Sadka (2008)).
In particular, the coefficients using levels of these variables are positive
and decay very slowly. However, when we look at changes (order imbalance)
and percentage changes (volume, volatility, and spreads) in these variables,
many display patterns that look similar to the return patterns shown in
Figure 1.

\noindent \textit{A. Volume and Order Imbalance}

To address trading-demand explanations of periodicity, we repeat the
cross-sectional regression using volume data,%
\begin{equation}
v_{i,t}=a_{k,t}+g_{k,t}v_{i,t-k}+u_{i,t},  \label{Vol}
\end{equation}%
where $v_{i,t}$ is the percentage change in the volume of stock $i$ traded
over the half-hour interval $t$. The means and $t$-statistics$^{9}$ of the
time series of volume responses are plotted in Panel A of Figure 5. The
pattern for $\widehat{g}_{t}$ strongly resembles the return response
pattern, $\widehat{\gamma }_{t}$, in Figure 1. In particular, the
cross-sectional volume response effects are uniformly negative at all lags
except multiples of 13. Figure 5 shows the pattern for 65 lagged half-hour
intervals corresponding to one week of calendar lags. Like the pattern of
return responses, the effect of volume responses lasts much longer. The
Internet Appendix (Figure IA.1) shows that the strength of daily volume
responses decays with longer lags, but remains positive and statistically
significant for up to 520 half-hour lags, corresponding to 40 days. Together
Figures 1 and 5 (and the longer-term results in Figure 2 and the Internet
Appendix) show that the intraday cross-sections of daily return and volume
display similar persistence lasting at least two months. Note that to the
extent that volume and volatility are correlated, the volume pattern is
consistent with the patterns in intraday volatility documented in Andersen
and Bollerslev (1997).

\begin{center}
[Figure 5 here]
\end{center}

It is probable that orders of different size convey different information,
and have different execution costs. Small trades are primarily generated by
individual investors, whereas large trades are preponderantly institutional
(Campbell, Ramadorai, and Schwartz (2009)). To check for differential
effects of volume generated by orders of different sizes, for each 30-minute
interval we determine the volume generated by trades of less than 1,000
shares each (small trade volume) and the volume generated by trades of
greater than or equal to 1,000 shares each (large trade volume). The
Internet Appendix (Figure IA.2) shows the volume responses separately for
small trades (Panel A) and large trades (Panel B). Percentage changes in
both small- and large-trade volume exhibit the same pattern of periodicity
that we see in returns in Figure 1 and in overall volume in Figure 5.

It is possible that institutions disguise large trades by breaking them into
smaller ones, making it difficult to detect their trades. Regardless of our
ability to identify specific types of trades, price pressure is presumably
caused by one-sided volume, not by balanced trading. Therefore, we also look
at the behavior of changes in signed volume, order imbalance. Order
imbalance is defined as the net signed volume over the interval, where the
sign is determined by a variant of the Lee and Ready (1991) trade
classification algorithm.$^{10}$ For example, if a time interval had 100,000
shares transacted that were classified as buyer-initiated and 75,000 shares
transacted that were classified as seller-initiated, then the order
imbalance for the period would be 25,000 shares (equal to 100,000 - 75,000).
Panel D of Figure 5 shows that changes in order imbalance exhibit
periodicity similar to but less pronounced than the periodicity in returns
and volume.

The Internet Appendix (Figure IA.2) indicates that order imbalance does not
exhibit any additional periodicity when partitioned into small versus large
trades, in contrast to the momentum results of Hvidkjaer (2008). Indeed, the
periodic pattern is absent in the order imbalance for large trades. Perhaps
trade classification algorithms for identifying buyer-initiated versus
seller-initiated trades produce noisy estimates for individual stocks over
short horizons such as the half-hour intervals used here.

\noindent \textit{B. Volatility and Bid-Ask Spreads}

Volatility tends to be high at the beginning and end of the trading day. In
addition, Andersen and Bollerslev (1997) find high returns for the S\&P 500
composite stock index futures contract at the beginning and end of the day.
It might be the case that movements in volatility are driving the return
periodicity we observe here.

Movements in bid-ask spreads might also be related to movements in volume,
volatility, and returns. For example, Admati and Pfleiderer (1988) derive a
model in which traders who have discretion on when to trade prefer to
coordinate trading at the same time. In this model, movements in volume,
volatility, and liquidity are related, in equilibrium.

Figure 5 (Panel B) shows that percentage changes in volatility (measured by
the absolute value of returns) exhibit intraday periodicity similar to that
found for returns and volume.

If market activity in individual stocks has daily periodicity, then bid-ask
spreads might follow the same pattern. We examine this to distinguish
changes in execution costs from changes in underlying returns. In addition,
since the high frequency return bias discussed in Blume and Stambaugh (1983)
is related to bid-ask bounce, we wish to test whether the return patterns we
find are related to systematic changes in bid-ask spreads. Panel C of Figure
5 shows the $t$-statistics from cross-sectional regressions in which
percentage changes in spreads are regressed on their lagged values. Spreads
do not exhibit the type of periodicity that we observe in returns.
Therefore, it appears that the return periodicity is not merely an artifact
of a spread-induced bias in estimated returns.

\noindent \textit{C. Does Periodicity in Related Variables Explain
Periodicity in Returns?}

We have shown that volume, order imbalance, and volatility have
cross-sectional predictability with the same daily pattern as returns,
whereas spreads do not. We now turn to the question of whether the
periodicity in these variables explain the return predictability we observe.
For example, daily patterns in institutional volume might affect stock
liquidity and volatility at the same time each day. News cycles might
trigger volume that causes investors to demand return premiums on a daily
cycle.

To control for these additional variables, we run cross-sectional
regressions of returns on lagged returns, including these additional lagged
regressors. The regression is%
\begin{equation}
r_{i,t}=\alpha _{k,t}+\gamma _{k,t}r_{i,t-k}+\delta _{k,t}^{\prime
}V_{i,t-k}+e_{i,t},
\end{equation}%
where the vector $V_{i,t-k}$ includes percentage changes in volume (total
shares traded during the lagged half-hour interval),%
\textbf{\ }volatility (absolute return), and relative bid-ask spread; and
absolute changes in order imbalance. If return predictability is caused by
predictable market activity based on these variables, then including them
should diminish or even subsume the predictability of returns based on past
returns. The values of $\widehat{\gamma }_{k}$ and the associated $t$%
-statistics are reported in Figure 6.

\begin{center}
[Figure 6 here]
\end{center}

Surprisingly, Figure 6 shows that the daily pattern of cross-sectional
return predictability is unaffected by including the additional regressors.
In other words, volume, order imbalance, and volatility have similar
patterns of intraday predictability, but these variables do not explain
daily return predictability.

As an additional refinement, we also regress returns on lagged returns,
lagged small-trade volume, and lagged large-trade volume. The Internet
Appendix (Figure IA.2) reports the results. The results show that the return
periodicity is not subsumed by large- or small-volume either.

\noindent \textit{D. Pre-decimalization Results}

We focus our analysis in this paper on the post-decimalization period. In
Table VIII we compare the return spreads on daily-frequency strategies (lags
13, 26, 39, 52, and 65) over our sample period to two additional sample
periods corresponding to tick sizes of one-eighth and one-sixteenth of a
dollar. We find that the statistical significance of the intraday
periodicity is greatest in the post-decimalization period. This might be due
to the increased use of trading algorithms (of the sort discussed above) by
institutional investors. Results during the period when the tick size is
one-sixteenth are quite unreliable and, hence, we present additional results
in Table VIII, Panel B, where the returns are Winsorized at the 1\% level
each month. That is, for each calendar month and each strategy (e.g., lag
13) we set all observations below (above) the $1^{st}$ ($99^{th}$)
percentile to the value of the $1^{st}$ ($99^{th}$) percentile. The
Winsorized results do not represent implementable trading strategies, but
reduce the influence of outliers. The Winsorized results with one sixteenth
tick size resemble the post-decimalization results, with slightly larger
coefficients and slightly smaller $t$-statistics. The larger coefficients
for the 1997 to 2000 period using the Winsorized estimator are not
consistent with the hypothesis that the increase in algorithmic trading is
driving the periodicity. The Winsorized returns over the 1993 to 1997
period, when the minimum price increment was one-eighth, have the smallest
coefficients and $t$-statistics.

\begin{center}
[Table VIII here]

\textbf{IV. Additional Diagnostics}
\end{center}

The previous sections demonstrate a pattern of daily return continuation.
This pattern is concentrated at the open and close, but also persists
throughout the day and lasts for many days. Because of the novelty of this
effect, it is important to verify that we have properly characterized the
pattern (pervasive and not restricted to special days or periods). This
section examines the pattern at a higher frequency and for different
calendar periods.

\noindent \textit{A. Five-Minute Interval Returns}

Figure 1 shows that when using half-hour intervals, sometimes there appears
to be slight spillover from one time interval to adjacent intervals. While
the return responses are most positive at exact multiples of one trading day
(e.g., 13-half-hour lags), they are often slightly positive at adjacent
intervals too (e.g., 12 or 14 half-hour lags). In this sense the choice of
half-hour intervals seems adequate.

It is conceivable, however, that our choice of interval masks higher
frequency phenomena. In addition, we would like to ensure that our results
are not sensitive to the exact methodological choices. Therefore, we
recalculate returns over five-minute intervals and estimate the return
responses of equation (1). In this case there are 78 five-minute intervals
during a trading day (omitting overnight returns).

Figure 7 shows these high-frequency results. Figure 7 looks like a noisy
version of Figure 1. The negative return reversal is most pronounced for the
first few five-minute lags, but persists for several days. And while there
is a clear daily effect at daily multiples of 78 lags, the leakage of the
neighboring five-minute intervals is also apparent. Overall, this graph
confirms the previous results and indicates that the half-hour return
intervals are fine enough to faithfully characterize high frequency returns.

\begin{center}
[Figure 7 here]
\end{center}

\noindent \textit{B. Day of Week}

A potential concern is the existence day-of-the-week effects. French (1980)
finds that the stock market earns different average returns on different
days of the week. In particular, average returns on the day following a
weekend are lower than average returns on other days. Therefore, we check
whether our daily effect is driven by a weekend effect or part of some other
weekly pattern.

The Internet Appendix (Table IA.VIII) shows the performance of our daily and
nondaily decile spread strategies on different days of the week. The effect
is remarkably consistent throughout the week.\textbf{\ }The Day 1 nondaily
strategy loses money at the open, mid-day, and close on every day. The
amounts range from -3.51 basis points on Thursdays to -6.14 basis points on
Mondays. Meanwhile the Day 1 daily strategy earns a positive premium when
averaged over all half-hour intervals. This ranges from 2.62 basis points on
Tuesdays to 3.44 basis points on Wednesdays. The results for longer lags are
weaker, but they have the same sign and are usually statistically
significant at the 5\% level. We conclude that the daily results are not
limited to a weekend effect or other weekday pattern.

\noindent \textit{C. Calendar Month and Turn-of-Month}

There are well-known monthly patterns in the stock market. This includes
both market-wide January effects and year-long seasonality (Rozeff and
Kinney (1976), Bouman and Jacobsen (2002), and Kamstra, Kramer, and Levi
(2003)) and cross-sectional return patterns such as the size effect at the
turn of the year (Keim (1983)). In addition to ruling out weekday effects,
we want to ensure that the daily pattern is not an artifact of some other
monthly seasonal phenomenon.

The Internet Appendix (Table IA.IX) reports the decile spread strategies for
every calendar month. Again, the results are strikingly consistent. The Day
1 nondaily strategy loses money in every calendar month, ranging from -2.88
basis points in June to -5.90 basis points in March. And the Day 1 daily
strategy makes money in every month, ranging from 1.70 basis points in March
to 7.80 basis points in November. The longer lag strategies have a similar
pattern, albeit smaller and less consistent. Nevertheless, the Day 5 daily
strategy is still profitable in every calendar month. The results are not
limited to a particular time of year, and certainly not limited to the turn
of the year.

While the results are not limited to turn of the year seasonality, they
might be driven by intra-month patterns. Ariel (1987) shows that stocks earn
a premium near the beginning and end of calendar months.

The Internet Appendix (Table IA.X) controls for turn-of-month effects by
separately reporting the combined results for trading days that occur on the
first or last day of the month. The results are remarkably consistent. The
Day 1 and Day 2 nondaily strategies lose money at the open, mid-day, and
close at both the turn of the month and the middle of the month, while the
daily strategies make money at all times. With few exceptions the Day 3, 4,
and 5 strategies maintain this pattern. The pattern is thus not related to
turn-of-the-month effects.

\noindent \textit{D. Index Membership}

If stocks rise and fall at the same time of day, then presumably there are
buyers and sellers who persistently trade them at those times (with
persistence in the direction of the trade). Index funds and benchmarked
mutual funds are natural suspects for these actions. These funds may have
large daily inflows or outflows and have an inelastic demand to invest those
funds to replicate the index. To economize on trading activity they might
perform \textquotedblleft basket trades\textquotedblright\ at the open of
trade, and to minimize tracking error they would have a motivation to trade
near the close. This is consistent with previous results showing a strong
effect at these times.

As a very rough cut at dealing with index fund behavior, we divide the
sample between firms in the S\&P 500 and those that are not. The Internet
Appendix (Table IA.XI) separates the decile spread results for S\&P 500
firms and non-S\&P 500 firms. The results for both daily and non-daily
strategies are generally stronger among the non-S\&P 500 stocks. For
example, the average decile spread based on the previous nondaily returns
loses 5.58 basis points in the non-S\&P 500 stocks, but loses less than one
basis point in the S\&P 500 index stocks. The Day 1 daily strategy earns
3.28 basis points with the non-S\&P 500 stocks, but earns only 2.19 basis
points with the index stocks. This difference masks some interesting
variation across the day. The closing-period daily effect is much larger,
relative to the intraday effect, for the S\&P 500 stocks. For example, the
ratio of the last period to intraday daily returns for Day 1 is 9.61
(9.42/0.98) for S\&P 500 stocks and 4.94 for non-S\&P 500 stocks. The daily
effect is larger for S\&P 500 stocks in absolute terms for Days 1 through 3,
and in relative terms (versus the mid-day period) for all five days. For the
opening half-hour the daily effects are smaller in absolute terms but larger
in relative terms for S\&P stocks. The large relative daily effect in the
closing-period for S\&P 500 stocks is consistent with index fund trading at
the close driving some of the daily effect. Similar effects are shown for
large versus small firms (reported above in Table IV), so it is not clear
whether this effect is an index effect or a large capitalization effect.

\begin{center}
\textbf{V. Conclusion}
\end{center}

A growing literature suggests that institutional investment flows influence
asset prices (e.g., Harris and Gruel (1986), Coval and Stafford (2007),
Boyer (2008), He and Krishnamurthy (2008), Lou (2008), Vayanos and Woolley
(2008)). We postulate that systematic trading and institutional fund flows
lead to predictable patterns, not only in trading volume and order
imbalances, but also in returns of common stocks. We study the periodicity
of cross-sectional differences in returns using half-hour observation
intervals in the period from January 2001 through December 2005. We document
pronounced intraday return reversals due to bid-ask bounce, and these
reversals last for several trading days. The market recovers from shocks to
depth in less than 60 minutes. However, we find significant continuation of
returns at intervals that are multiples of a day and this effect lasts for
at least 40 trading days. This daily periodicity is of the same order of
magnitude as current institutional commission rates and the quoted
half-spread.

Changes in trading volume, order imbalances, and volatility exhibit similar
patterns, but do not explain the return patterns. The return continuation at
daily frequencies is more pronounced for the first and last half-hour
periods. These effects are not driven by firm size, systematic risk premia,
or inclusion in the S\&P 500 index (as a proxy for trading by index funds).
The pattern is also not driven by particular months of the year, days of the
week, or by turn-of-the-month effects. The periodicity at the daily
frequency is observed when we also use bid-to-bid, ask-to-ask, or
midpoint-to-midpoint returns, which implies that the periodicity is not
merely due to patterns of where transactions prices occur relative to bid
and ask prices.

The results are consistent with investors having a predictable demand for
immediacy at certain times of the day. The pattern does not present a profit
opportunity in the absence of other motives to trade, since strategies that
attempt to take advantage of the daily periodicity lose money, after paying
the bid-ask spread. However, the magnitude of the return pattern is sizeable
relative to several components of transactions costs such as commissions and
effective spreads.\pagebreak

\linewidth= 6.9in

\textbf{References}

\textbf{\ \vspace{-0.35in}}

\begin{quote}
\parindent=-26pt \hspace{-26pt} \baselineskip= 12pt

Admati, Anat, and Paul Pfleiderer, 1988, Theory of intra-day patterns:
Volume and price variability, \textit{Review of Financial Studies} 1, 3-40.

Almgren, Robert, and Neil Chriss, 2000, Optimal execution of portfolio
transactions, \textit{Journal of Risk }3, 5-39.

Almgren, Robert, and Julian Lorenz, 2006, Bayesian adaptive trading with a
daily cycle, \textit{Journal of Trading }1, 38-46.

Andersen, Torben G., and Tim Bollerslev, 1997, Intraday periodicity and
volatility persistence in financial markets, \textit{Journal of Empirical
Finance} 4, 115-158.

Ariel, Robert A., 1987, A monthly effect in stock returns, \textit{Journal
of Financial Economics} 18, 161-174.

Avramov, Doron, Tarun Chordia, and Amit Goyal, 2006, Liquidity and
autocorrelations in individual stock returns, \textit{Journal of Finance}
61, 2365-2394.

Bertsimas, Dimitris, and Andrew W. Lo, 1998, Optimal control of execution
costs, \textit{Journal of Financial Markets} 1, 1-50.

Blackburn, Douglas W., William N. Goetzmann, and Andrey D. Ukhov, 2007, Risk
aversion and clientele effects, Working paper, Indiana University.

Blume, Marshall E., and Robert F. Stambaugh, 1983, Biases in computed
returns: An application to the size effect, \textit{Journal of Financial
Economics} 12, 387-404.

Bollerslev, Tim, 1986, Generalized autoregressive conditional
heteroskedasticity, \textit{Journal of Econometrics} 31, 307-327.

Bouman, Sven, and Ben Jacobsen, 2002, The Halloween indicator, `Sell in May
and go away': Another puzzle, \textit{American Economic Review} 92,
1618-1635.

Boyer, Brian H., 2008, Comovement among stocks with similar book-to-market
ratios, Working paper, Brigham Young University.

Breen, William J., Laurie Simon Hodrick, and Robert A. Korajczyk, 2002,
Predicting equity liquidity, \textit{Management Science} 48\textbf{,}
470-483.

Campbell, John Y., Tarun Ramadorai, and Allie Schwartz, 2009, Caught on
tape: Institutional trading, stock returns, and earnings announcements, 
\textit{Journal of Financial Economics }92, 66--91.

Carhart, Mark M., Ron Kaniel, David K. Musto, and Adam V. Reed, 2002,
Leaning for the tape: Evidence of gaming behavior in equity mutual funds, 
\textit{Journal of Finance} 58, 661-693.

Chordia, Tarun, Richard Roll, and Avanidhar Subrahmanyam, 2005, Evidence on
the speed of convergence to market efficiency, \textit{Journal of Financial
Economics} 76, 271-292.

Coppejans, Mark, Ian Domowitz, and Ananth Madhavan, 2004, Resiliency in an
automated auction, Working paper, ITG, Inc.

Coval, Joshua, and Erik Stafford, 2007, Asset fire sales (and purchases) in
equity markets, \textit{Journal of Financial Economics} 86, 479--512.

Del Guercio, Diane, and Paula A. Tkac, 2002, The determinants of the flow of
funds of managed portfolios: Mutual funds vs. pension funds, \textit{Journal
of Financial and Quantitative Analysis} 37, 523-557.

Dimson, Elroy, 1979, Risk measurement when shares are subject to infrequent
trading, \textit{Journal of Financial Economics} 7, 197-226.

Engle, Robert, and Robert Ferstenberg, 2007, Execution risk: It's the same
as investment risk, \textit{Journal of Trading }2, 10-20.

Fama, Eugene F., 1976, \textit{Foundations of Finance} (Basic Books, New
York).

Fama Eugene F., and James D. MacBeth, 1973, Risk, return, and equilibrium:
Empirical tests, \textit{Journal of Political Economy} 81, 607-636.

Frazzini, Andrea, and Owen A. Lamont, 2008, Dumb money: Mutual fund flows
and the cross-section of stock returns, \textit{Journal of Financial
Economics} 88, 299--322.

French, Kenneth R., 1980, Stock returns and the weekend effect, \textit{%
Journal of Financial Economics} 8, 55-69.

Glosten, Lawrence R., and Lawrence E. Harris, 1988, Estimating the
components of the bid/ask spread,\ \textit{Journal of Financial Economics}
21, 123-142.

Glosten, Lawrence R., and Paul R. Milgrom, 1985, Bid, ask and transaction
prices in a specialist market with heterogeneously informed traders, \textit{%
Journal of Financial Economics} 14, 71-100.

Goyenko, Ruslan Y., Craig W. Holden, and Charles A. Trzcinka, 2009, Do
liquidity measures measure liquidity? \textit{Journal of Financial Economics}
92, 153-181.

Grinold, Richard C., and Ronald N. Kahn, 2000, \textit{Active} \textit{%
Portfolio Management: A Quantitative Approach for Producing Superior Returns
and Selecting Superior Returns and Controlling Risk, }2nd edition
(McGraw-Hill, New York).

Gurliacci, Mark, David Jeria, and George Sofianos, 2009, When the going gets
tough, the algos get going, \textit{Journal of Trading} 4, 34-44.

Harris, Lawrence, 1986, A transaction data study of weekly and intradaily
patterns in stock returns, \textit{Journal of Financial Economics} 16,
99-117.

Harris, Lawrence, and Eitan Gurel, 1986, Price and volume effects associated
with changes in the S\&P 500 list: New evidence for the existence of price
pressures, \textit{Journal of Finance} 41, 815--829.

Hasbrouck, Joel, 2009, Trading costs and returns for U.S. equities:
Estimating effective costs from daily data, \textit{Journal of Finance} 64,
1445-1477.

He, Zhiguo, and Arvind Krishnamurthy, 2008, Intermediary asset pricing,
Working paper, Northwestern University.

Hendershott, Terrence, Charles M. Jones, and Albert J. Menkveld, 2009, Does
algorithmic trading improve liquidity? Working paper, University of
California, Berkeley.

Heston, Steven L., and Ronnie Sadka, 2008a, Seasonality in the cross-section
of stock returns, \textit{Journal of Financial Economics} 87, 418-445.

Heston, Steven L., and Ronnie Sadka, 2008b, Seasonality in the cross-section
of stock returns: The international evidence, Forthcoming, \textit{Journal
of Financial and Quantitative Analysis}.

Holden, Craig W., and Avanidhar Subrahmanyam, 2002, News events, information
acquisition, and serial correlation, \textit{Journal of Business }75, 1-32.

Hora, Merrell, 2006, Tactical liquidity trading and intraday volume, Working
paper, Credit Suisse.

Huberman, Gur, and Werner Stanzl, 2005, Optimal Liquidity Trading, \textit{%
Review of Finance} 9, 165--200.

Hvidkjaer, Soeren, 2008, Small trades and the cross-section of stock
returns, \textit{Review of Financial Studies} 21, 1123-1151.

Jain, Prem C., and Gun-Ho Joh, 1988, The dependence between hourly prices
and trading volume, \textit{Journal of Financial and Quantitative Analysis}
23, 269-283.

Jegadeesh, Narasimhan, 1990, Evidence of predictable behavior of security
returns, \textit{Journal of Finance} 45, 881-898.

Jegadeesh, Narasimhan, and Sheridan Titman, 1993, Returns to buying winners
and selling losers: Implications for stock market efficiency. \textit{%
Journal of Finance} 48, 65-91.

Jeria, David, and George Sofianos, 2008, Algo evolution: From 4Cast to
OptimIS and lower shortfall, \textit{Street Smart} 37, 1-5.

Kamstra, Mark J., Lisa A. Kramer, and Maurice D. Levi, 2003, Winter blues:
Seasonal affective disorder (SAD) stock market returns, \textit{American
Economic Review} 93, 324-343.

Keim, Donald B., 1983, Size-related anomalies and stock return seasonality:
Further evidence, \textit{Journal of Financial Economics} 12, 13-32.

Keim, Donald B., 1989, Trading patterns, bid-ask spreads, and estimated
security returns: The case of common stocks at calendar turning points, 
\textit{Journal of Financial Economics} 25, 75-97.

Korajczyk, Robert A., and Ronnie Sadka, 2004, Are momentum profits robust to
trading costs? \textit{Journal of Finance} 59, 1039-1082.

Korajczyk, Robert A., and Ronnie Sadka, 2008, Pricing the commonality across
alternative measures of liquidity, \textit{Journal of Financial Economics}
87, 45-72.

Large, Jeremy, 2007, Measuring the resiliency of an electronic limit order
book, \textit{Journal of Financial Markets} 10, 1--25.

Lee, Charles M. C., and Mark J. Ready, 1991, Inferring trade direction from
intraday data, \textit{Journal of Finance} 46, 733-754.

Lehmann, Bruce, 1990, Fads, martingales and market efficiency, \textit{%
Quarterly Journal of Economics} 105, 1-28.

Lehmann, Bruce N., and David M. Modest, 2005, Diversification and the
optimal construction of basis portfolios, \textit{Management Science }51,
581--598.

Lo, Andrew W., and A. Craig MacKinlay, 1990, When are contrarian profits due
to stock market overreaction? \textit{Review of Financial Studies} 3,
175-208.

Lo, Andrew W., and Jiang Wang, 2006, Trading volume: Implications of an
intertemporal capital asset pricing model, \textit{Journal of Finance} 61,
2805-2840.

Lou, Dong, 2008, A flow-based explanation for return predictability, Working
paper, Yale University.

Pagano, Michael S., Lin Peng, and Robert A. Schwartz, 2008, The quality of
price formation at market openings and closings: Evidence from the Nasdaq
stock market, Working paper, Villanova University.

Roll, Richard, 1984, A simple implicit measure of the effective bid-ask
spread in an efficient market, \textit{Journal of Finance} 39, 1127-1139.

Rozeff, Michael S., and William R. Kinney, 1976, Capital market seasonality:
The case of stock returns, \textit{Journal of Financial Economics} 3,
379-402.

Sadka, Ronnie, 2006, Momentum and post-earnings-announcement drift
anomalies: The role of liquidity risk, \textit{Journal of Financial Economics%
} 80, 309--349.

Stoll, Hans R., 1978, The supply of dealer services in securities markets, 
\textit{Journal of Finance} 33, 1133-1151.

Tversky, Amos, and Daniel Kahneman, 1974, Judgment under uncertainty:
Heuristics and biases, \textit{Science} 185, 1124-1131.

Vayanos, Dimitri, 2001, Strategic trading in a dynamic noisy market, \textit{%
Journal of Finance} 56, 131-171.

Vayanos, Dimitri, and Paul Woolley, 2008, An institutional theory of
momentum and reversal, Working paper, London School of Economics.

Wood, Robert A., Thomas H. McInish, and J. Keith Ord, 1985, An investigation
of transactions data for NYSE stocks, \textit{Journal of Finance} 40,
723-739.

\end{quote}


\pagebreak \linewidth= 6.9in

\textbf{Footnotes}

\textbf{\ \vspace{-0.35in}}

\begin{quote}
\parindent=-26pt \hspace{-26pt} \baselineskip= 12pt
\end{quote}

1.\qquad Following Fama (1976), these responses have the interpretation of
excess returns on costless portfolios that have excess returns of 100\% in a
previous half-hour interval.

2.\qquad We use Fama and MacBeth (1973) $t$-statistics, which assume no
autocorrelation in the coefficients of the cross-sectional regressions,
because we find no evidence of significant autocorrelation in the estimates, 
$\widehat{\gamma }_{k}$.

3.\qquad The Internet Appendix is available on the \textit{Journal of Finance%
} web site, at \newline
http://www.afajof.org/supplements.asp.

4.\qquad The returns on the separate winners and losers portfolios are
available in the Internet Appendix (Table IA.II).

5.\qquad An 8\% equity risk premium corresponds to approximately 3.2 basis
points per trading day (that is, 800/250).

6.\qquad Korajczyk and Sadka (2004), Goyenko, Holden, and Trzcinka (2009),
and Hasbrouck (2009) find larger quoted spreads. \ However, these papers
have sample periods that include both pre-decimalization data and some
include less liquid stocks.

7.\qquad In the Internet Appendix (Table IA.XI) we estimate the return
spread for S\&P 500 stocks (which is the sample studied in Gurliacci, et
al.) to be 2.19 basis points.

8.\qquad In the Internet Appendix (Figure IA.3) we present a version of
Figure 4 that uses different scaling of the return results, which makes it
easier to discern the return patterns for the mid-day period.

9.\qquad The estimates, $\widehat{g}_{t,k}$, do not display substantial
autocorrelation so we do not make adjustments to the Fama-MacBeth standard
errors.

10.\qquad Given the increased transactions volume in our sample period
relative to that in Lee and Ready (1991), we do not require their five
second minimum period between quotes and transactions.

\end{document}